\pdfoutput=1
\documentclass[aps,pra,twocolumn,10pt]{revtex4-1}
\usepackage[utf8]{inputenc}
\usepackage{fourier,mathtools,graphicx,microtype}
\usepackage[cal=boondoxo]{mathalfa}
\usepackage[colorlinks=true,citecolor=blue,urlcolor=blue]{hyperref}

\begin{document}

\title{Radiative Decay of Bound Electron Pairs in Two-Dimensional Topological Insulators}

\author{Vladimir A.\ Sablikov\textsuperscript{1,2} and Bagun S.\ Shchamkhalova\textsuperscript{2}}
\affiliation{\textsuperscript{1} Kotelnikov Institute of Radio Engineering and Electronics of Russian Academy of Sciences.  Mokhovaya 11, bld.7, 125009 Moscow, Russia \\
\textsuperscript{2} Fryazino Branch of the Kotelnikov Institute of Radio Engineering and Electronics, Russian Academy of Sciences, Fryazino, Moscow District, 141190, Russia}


\begin{abstract}
Bound electron pairs (BEPs) with energy in the band gap are interesting because they can participate in charge and spin transport in modern topologically nontrivial materials. We address the problem of their stability and study the radiative decay of the BEPs formed due to the negative reduced effective mass in two-dimensional topological insulators. The decay time is found to be rather large on the scale of the characteristic relaxation times of the electron system and significantly dependent on the topological properties and dispersion of the band states. In topological phase the decay time is much longer than in the trivial one, and is estimated as $\sim$1~ns for the HgTe/CdHgTe heterostructures. However, the longest decay time is in the topological phase with nearly flat dispersion in the band extema.
\end{abstract}

\maketitle

\section{Introduction}

Bound electron pairs (BEPs) are charged composite bosons that arise in crystalline solids despite of the Coulomb repulsion of the electrons. The great interest in BEPs is associated not only with the problems of superconductivity~\cite{combescot2015excitons,kagan2013modern,RevModPhys.62.113}, but also the effects of Coulomb interaction in modern topologically nontrivial materials~\cite{Rachel_2018}. Recent studies have revealed new mechanisms for the Coulomb pairing, which lead to the formation of the BEPs with unusual and yet insufficiently studied properties~\cite{doi:10.1002/pssb.201800584,PhysRevB.92.085409,PhysRevB.95.085417,PhysRevB.98.115137}. An important place among them belongs to the mechanism caused by a negative reduced mass of electrons, since it can lead to the formation of the BEPs with a sufficiently large binding energy. 

The basic idea of this mechanism was proposed in Ref.~\cite{gross1971inverse} for BiI\textsubscript{3}. Further studies have shown that it is rather universal in nature and can be generalized for a wide variety of materials. The BEPs of this nature were studied for ordinary crystals~\cite{MahajanJPhysA2006,SouzaClaroPRB2010,HaiJPhysCM2014}, graphene~\cite{downing2017bielectron,doi:10.1002/pssb.201800584}, bigraphene~\cite{PhysRevB.92.085409}, topological insulators~\cite{PhysRevB.95.085417}, and it was found that the properties of the BEPs are very different for different systems. Of great interest are topological insulators, in which strong mixing of the electron and hole states significantly contributes to the pairing of electrons giving rise to the formation of BEPs with a higher binding energy, which is comparable with the band gap~\cite{PhysRevB.95.085417}. The pairing of electrons due to the mechanism of the negative reduced mass was considered as one of the mechanisms of high-temperature superconductivity~\cite{Belyavskii2000}. The BEPs of the similar nature are formed also by cold atoms in optical lattices, where they are called doublons~\cite{winkler2006repulsively,PhysRevB.89.195119,Han_2016}.

Stable BEPs with high binding energy and the energy in the band gap are of great interest because they can efficiently transfer charge and spin, and their interaction can lead to nontrivial collective effects. A sufficiently large number of the BEPs to produce observable effects may be present in equilibrium, but the most promising method for their generation and manipulation is optical excitation. It is obvious that in this case the number of the BEPs is determined by their decay time. However, the decay of the BEPs, as far as we know, has not yet been studied, although researches in this direction are developing for doublons. So, the dynamics and decay of doublons was recently studied in the Fermi-Hubbard model~\cite{PhysRevLett.104.080401,PhysRevB.85.205127,PhysRevB.96.045408}.

It is clear that the BEPs can decay with the emission of photons or phonons or both, but the decay of BEPs differs substantially from that of excitons, which is widely studied in the literature~\cite{PhysRevB.38.1228,ANDREANI1991641,PhysRevB.93.205423,RevModPhys.90.021001}. First, the fact is important that the decay of a BEP occurs with appearance of two free particles, due to which the conditions for the conservation of energy, momentum and spin are significantly different from those for excitons. Second, in topologically nontrival materials the Hamiltonian of the interaction of electrons with the light differs from usual dipole Hamiltonian in the trivial case. Third, the two-particle wave functions are represented by high-rank spinors, whose components describe various combinations of spins and pseudospins of paired electrons. Moreover, the components have different and rather complicated spatial distribution.

In this article, we study the radiative decay of BEPs in two-dimensional materials with a two-band spectrum described by the Bernevig-Hughes-Zhang (BHZ) model~\cite{Bernevig1757}, which is applicable to both the topological and trivial phases. The phonon mechanism of the BEP decay is beyond the scope of this paper. Recently, we have found that a wide set of two-electron bound states with different spin structure and atomic orbital configurations can exist in these materials~\cite{PhysRevB.95.085417}. Here, we obtain the two-particle Hamiltonian of the electron interaction with light, analyze possible decay channels and calculate the decay rate for a variety of the BEP types in the topological and trivial phases.

\section{Two-electron states}

To begin, we briefly describe the classification and properties of two-electron states in the frame of the BHZ model~\cite{Bernevig1757}. The two-band BHZ Hamiltonian is formed by a $s$-type band ($\Gamma_6$) and a $p$-type band split by spin-orbit interaction into a $J$=3/2 band ($\Gamma_8$) and a $J$=1/2 band. The electron ($e$-) states are formed by the $\Gamma_6$ band and the light-hole $\Gamma_8$ sub-band, and the hole ($h$-) states are formed by the heavy-hole $\Gamma_8$ sub-band. The $e$- and $h$-states have a definite spin projection $s_z=\pm 1$, so the single-particle basis is $\left(|e\uparrow\rangle,|h\uparrow\rangle,|e\downarrow\rangle,|h\downarrow\rangle\right)^T$.

An important parameter of the model is the value $a = A/\sqrt{|MB|}$, which characterizes the hybridization of the $e$- and $h$-bands. Here $A$, $M$ and $B$ are the parameters of the BHZ model: $M$ is the mass term, $B$ is the parameter of the dispersion in the $e$- and $h$-bands, which are assumed to be symmetric, $A$ describes the hybridization of the $e$- and $h$-bands. In the topological phase, the $e$- and $h$-bands are inverted and $MB>0$. The parameter $a$ determines the dispersion in the conduction ($c$-) and valence ($v$-) bands formed as a result of the hybridization of the $e$- and $h$-bands. For $a>2$, the band dispersion is quadratic near the extrema with positive effective mass at the $c$-band bottom. For $a=\sqrt{2}$, the band dispersion is nearly flat at the extrema of the bands (the dispersion has the form $E\sim\pm\sqrt{M^2+B^2 k^4}$, and effective mass is infinite when $k\to 0$. For $a<\sqrt{2}$, the dispersion has a mexican-hat shape).

\subsection{Free electrons}

The wave functions of the single-electron states $\psi_{s,\mathbf{k}}^{\lambda}(\mathbf{r})$ are characterized by the band index $\lambda=\pm 1$, which indicates $c$- and $v$-bands, the spin projection $s$ and wave vector $\mathbf{k}$. The single-particle energy is degenerate with respect to the spin 
\begin{equation}
\varepsilon_{s,\mathbf{k}}^{\lambda}=\delta k^2 + \lambda \sqrt{(-\mu +k^2)^2+a^2k^2}\,.
\end{equation}
Hereinafter we use dimensionless quantities: the energy is normalized to $|M|$ and the wave vector $\mathbf{k}$ is normalized to $\sqrt{|M/B|}$. $\delta$ is the parameter of the asymmetry of the $e$- and $h$-bands in the BHZ model, that will be assumed to be zero in specific calculations below. $\mu=\mathrm{sign}(MB)$ stands to define the topological and trivial phases.

The two-particle wave function of free electrons, antisymmetric with respect to the permutation of the particles, reads 
\begin{multline}
\Phi_{s_1,\mathbf{k_1};s_2,\mathbf{k_2}}^{\lambda_1,\lambda_2}(\mathbf{r}_1,\mathbf{r}_2)=\frac{1}{2}\left[\psi_{s_1,\mathbf{k}_1}^{\lambda_1}(\mathbf{r}_1)\otimes\psi_{s_2,\mathbf{k}_2}^{\lambda_2}(\mathbf{r}_2) \right. \\ \left. - \psi_{s_2,\mathbf{k}_2}^{\lambda_2}(\mathbf{r}_1)\otimes\psi_{s_1,\mathbf{k}_1}^{\lambda_1}(\mathbf{r}_2)\right]\,
\end{multline}
where the single-particle wave functions are:
\begin{equation}
\psi_{\uparrow,\mathbf{k}}^{\lambda}=C_{\mathbf{k}}^{\lambda}
\begin{pmatrix}
1 \\ g_{\uparrow,\mathbf{k}}^{\lambda}\\0\\0
\end{pmatrix}
 e^{i\mathbf{k}\mathbf{r}},\;\;
\psi_{\downarrow,\mathbf{k}}^{\lambda}=C_{\mathbf{k}}^{\lambda}
\begin{pmatrix}
0\\0\\1 \\ g_{\downarrow,\mathbf{k}}^{\lambda}
\end{pmatrix}
 e^{i\mathbf{k}\mathbf{r}},
\end{equation}
with 
\begin{equation}
g_{\uparrow,\mathbf{k}}^{\lambda}=\frac{a (k_x-i k_y)}{\varepsilon_{\lambda}-\mu+k^2},\;g_{\downarrow,\mathbf{k}}^{\lambda}=\frac{-a (k_x+i k_y)}{\varepsilon_{\lambda}-\mu+k^2}.
\end{equation}

Spin of the two-electron states is determined only by the spin projection $S_z$, since $S_z$ is the only conserving spin quantity and $S^2$ does non conserve in the BHZ model. Therefore, the two-electron states are classified as the singlets, with the spins being opposite, and the triplets with parallel spins. Band indices $\lambda_{1,2}$ define the composition of the atomic orbitals (the Bloch functions of the $e$- and $h$-bands) that form a given two-electron states. It is clear that there are only three types of the states in which the index set $(\lambda_1, \lambda_2)$ is equal to $(c, c)$, $(v, v)$ and $(c, v)$.

The two-particle energy spectrum
\begin{equation}
\varepsilon_{s_1,\mathbf{k_1};s_2,\mathbf{k_2}}^{\lambda_1,\lambda_2}=\varepsilon_{s_1,\mathbf{k}_1}^{\lambda_1}+\varepsilon_{s_2,\mathbf{k}_2}^{\lambda_2}
\end{equation}
contains two bands of propagating states with energy $|\varepsilon|>2\varepsilon_g$, where $\varepsilon_g$ is the gap in the single-particle spectrum, Fig~1(a). In these bands, both electrons are in the same ($c$- or $v$-) band. There is another branch of the propagating states, which are formed by the electrons in different bands. We will be interested in two-electron states with an almost zero total momentum, since the wave vector of the photons is much smaller than that of the electrons. In the general case, where $e$- or $h$-bands are asymmetric, this branch covers a wide energy region ($\varepsilon>0$ for $\delta>0$). In the symmetric case where $\delta=0$, this branch degenerates into the fixed energy $\varepsilon=0$.

\begin{figure}[t]%
\centerline{\includegraphics*[width=0.95\linewidth,height=0.56\linewidth]{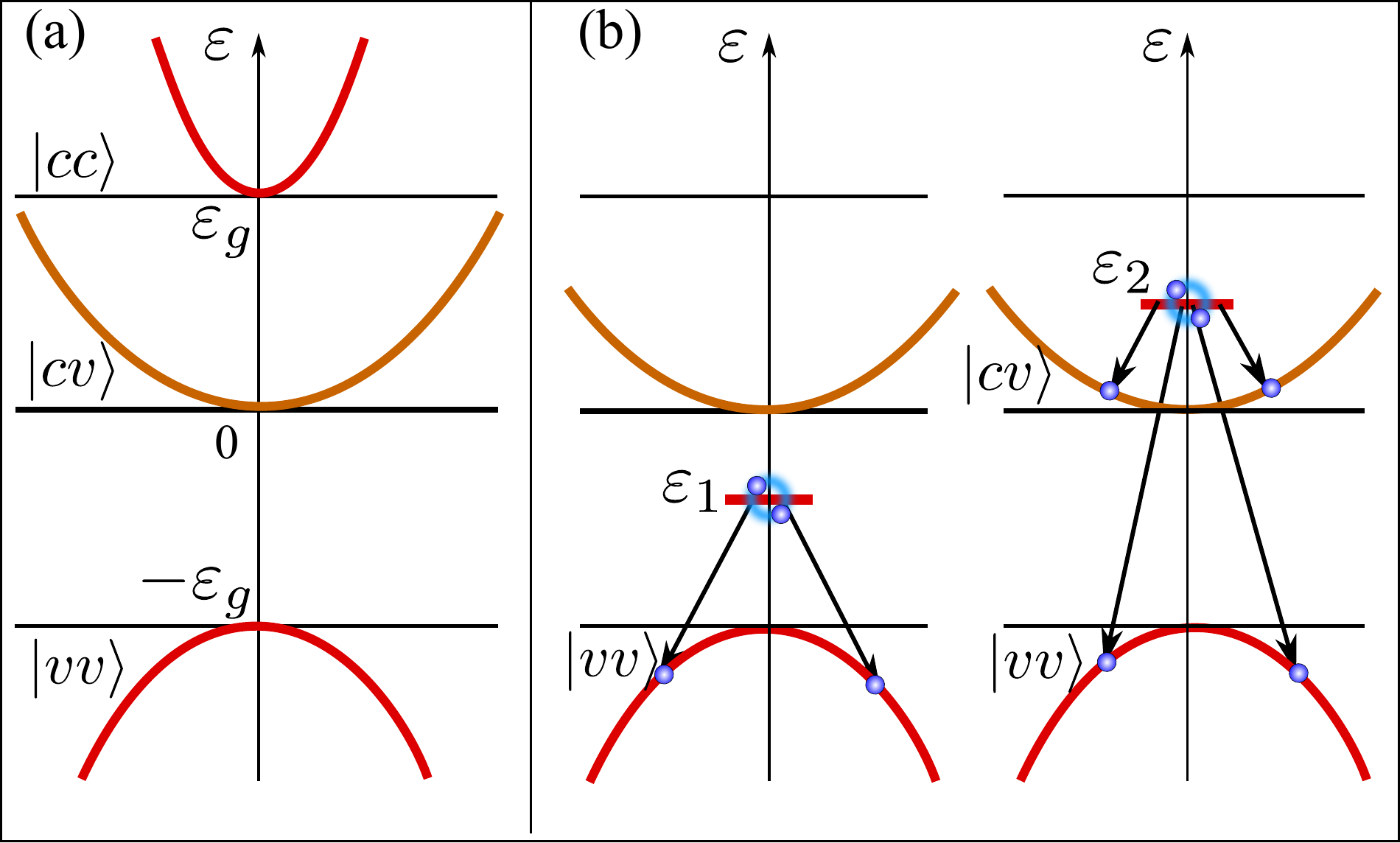}}
\caption{(a) Two-particle spectrum of free electrons. Thick lines show the band spectra $\varepsilon_{\mathbf{k}_1,\mathbf{k}_2}$ and the branch of the electrons from different bands. (b) Energy levels of the two-electron bound states of the first and second types and the electron transitions leading to the decay of the BEPs with the energy below and above the middle of the gap.}
\label{fig1}
\end{figure}

\subsection{Two-electron bound states}

The two-electron bound states were studied in detail in Ref.~\cite{PhysRevB.95.085417}. They appear in both the topological and trivial phases, when the electron-electron (\textit{e-e}) interaction potential $v$ is large enough. The bound state energy $\varepsilon_{bs}$ lies in the gap of the band spectrum. 

Spin state of the BEPs is characterized by the spin projection $S_z$, which can be zero (singlet state) and $\pm 1$ (two triplet states), quite similarly to the free states. However, there is no conserving quantity to characterize the pseudospin of the bound states. Nevertheless, when the \textit{e-e} interaction is not strong, the BEPs can be divided into two groups with quantitatively different pseudospin composition. In the states of the first type, both paired electrons are mainly in the $e$-band, i.e. the basis components $|e\uparrow\rangle$ and $|e\downarrow\rangle$ prevail in the wave function. The energy of these states is close to the $v$-band, Fig.~\ref{fig1}(b). It the second-type states, one electron is mainly in the $e$-band and the other is in the $h$-band. The energy level of these states lies above the middle of the gap.

\begin{figure}[t]%
\centerline{\includegraphics*[width=1.0\linewidth,height=0.6\linewidth]{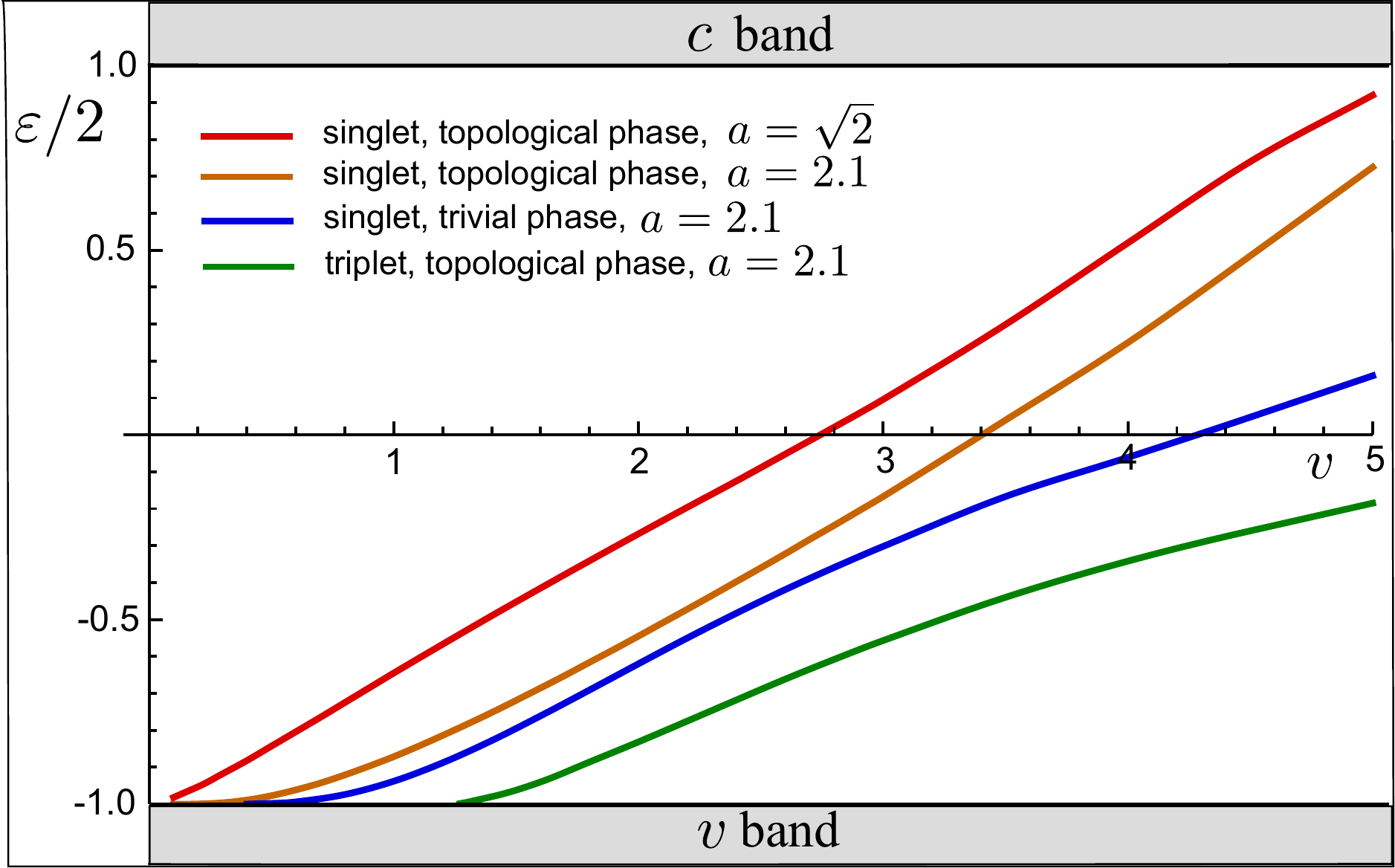}}
\caption{The ground state energy of two-electron bound states of the first type as a function of the \textit{e-e} interaction potential in the topological and trivial phases with different band spectrum. $v$ is normalized to $|M|$.}
\label{fig2}
\end{figure}

In what follows we focus on the BEPs with largest binding energy, since they are expected to be most stable. The states of the second group are less stable because in the general case (where $\delta \ne 0$) their energy lies in the continuum of the free states $\Phi_{s_1,\mathbf{k_1};s_2,\mathbf{k_2}}^{c,v}$. In addition, they can decay through two channels shown in Fig.~\ref{fig1}(b): the $vv$-channel, where both electrons pass into the $v$ band, and the $cv$-channel, where one electron passes into the $v$-band and other into the $c$-band. Therefore we consider the states of the first type. Their ground-state energy is shown in Fig.~\ref{fig2} as a function of the interaction potential amplitude $v$ for different cases. Shown are the singlet states for topological and trivial phases at the hybridization parameter $a>2$, the triplet state for the same $a$, and the singlet state in the case, where $a=\sqrt{2} $ and the band spectrum is nearly flat in the band edges. It is seen that the singlet states have a higher binding energy than the triplet state, all other things being equal. Note that the two-electron bound state arising at $a=\sqrt{2}$ have the highest binding energy. This type of the BEPs exists only in the topological phase. 

\begin{figure}[htb]%
\centerline{\includegraphics*[width=\linewidth,height=0.4\linewidth]{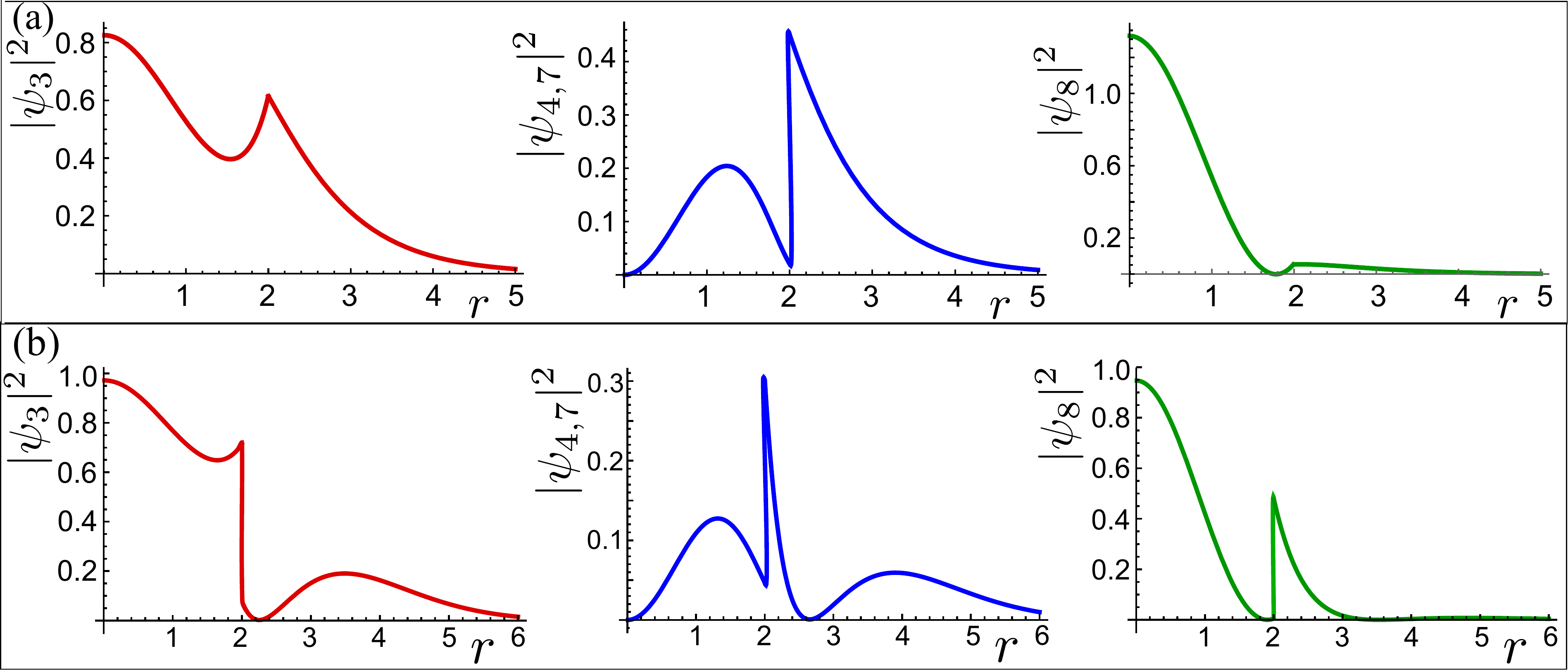}}
\caption{Spinor components of the singlet two-electron bound states with highest binding energy in the topological phase with (a) nearly quadratic band dispersion $a$=2.1, $v$=2.0, $r_0$=2.0, $\varepsilon /2$=-0.545, and (b) nearly flat band dispersion  $a=\sqrt{2}$, $v$=2.0, $r_0$=2.0, $\varepsilon /2$=-0.646.}
\label{fig3}
\end{figure}

The two-electron wave functions of the bound states are presented in the form of 16-rank spinor in the basis $\left\{\left(|e\uparrow\rangle,|h\uparrow\rangle,|e\downarrow\rangle,|h\downarrow\rangle\right)_1^T\otimes\left(|e\uparrow\rangle,|h\uparrow\rangle,|e\downarrow\rangle,|h\downarrow\rangle\right)_2^T\right\}$. For the singlet states they have the form~\cite{PhysRevB.95.085417}: 
\begin{multline}
\Psi_s(\mathbf{r})=\left(0, 0, \psi_3(r), \psi_4(r)e^{i\varphi}, 0, 0, \psi_7(r)e^{-i\varphi}, \psi_8(r),\right. \\ \left. -\psi_3(r), \psi_7(r)e^{-i\varphi}, 0, 0, \psi_4(r)e^{i\varphi}, \psi_8(r), 0, 0\right)^T,
\label{bound_state_spinor}
\end{multline}
where $\mathbf{r}=\mathbf{r}_1-\mathbf{r}_2$ is the vector of the relative position of the electrons and $\varphi$ is the angular coordinate. For simplicity, we consider the BEPs with zero total momentum. The independent components of the spinor~(\ref{bound_state_spinor}) as functions of the distance $r$ are illustrated in Fig.~\ref{fig3} for the cases of $a>2$ and $a=\sqrt{2}$. The spinor components are seen to have very different and unusual spatial distribution. The calculations were carried out for a step-like interaction potential of a radius $r_0$, which well approximates the short range interaction of electrons. The specific structure of the spinor and the coordinate dependence of its components plays an important role in the calculation of the decay probability. 

\section{Radiative decay probability}

The radiative decay of the BEPs is determined by two-particle Hamiltonian $H'(1,2)$ of their interaction with light. We obtain this Hamiltonian from the two-electron Hamiltonian of the BHZ model within of the electric dipole approximation by substitution $\mathbf{k}\to \mathbf{k}+(e/\hbar c)\mathbf{A}$, with $\mathbf{A}$ being a vector potential. In the dimensionless form the vector potential $\mathbf{A}$ describing the photon emission is
\begin{equation}
\mathbf{A}(\mathbf{r,t})=\sum_{\mathbf{q},\nu} \sqrt{\frac{2\pi e^2|B|}{V\kappa M^2 \varepsilon_{\mathbf{q}}}} \mathbf{e}_{\nu}^* a_{\mathbf{q},\nu}^{\dag}e^{i(\varepsilon_\mathbf{q}t-\mathbf{q}\mathbf{r})},
\end{equation} 
where $\kappa$ is the dielectric constant of the material, $V$ is normalization volume, $\varepsilon_q=\hbar\omega_q/|M|$ is photon energy, $\mathbf{q}$ is photon wave vector, and $\mathbf{e}_{\nu}$ is polarization vector.

The Hamiltonian of the interaction of the BEPs with light has the form:
\begin{multline}
H'(1,2)=4 \mathbf{A}\mathbf{k} \cdot \mathbf{M}_0 \\  +A_+(\mathbf{M}_+ \!\otimes \!\mathbf{I}_4 \!+\!\mathbf{I}_4 \!\otimes  \!\mathbf{M}_+) \!+ \! A_-(\mathbf{M}_- \!\otimes \! \mathbf{I}_4\!+\!\mathbf{I}_4 \!\otimes \! \mathbf{M}_-),
\label{light-pair}
\end{multline}
where $\mathbf{M}_0$ and $\mathbf{M}_{\pm}$ are numerical matrices:
\begin{gather}
\mathbf{M}_0=\mathrm{diag}[0, 1,0, 1, -1, 0, -1, 0, 0, 1,0, 1, -1, 0, -1, 0]\,,\\
\mathbf{M}_+=
\begin{pmatrix}
0&1&0&0\\0&0&0&0\\0&0&0&0\\0&0&-1&0
\end{pmatrix},\:
\mathbf{M}_-=
\begin{pmatrix}
0&0&0&0\\1&0&0&0\\0&0&0&-1\\0&0&0&0
\end{pmatrix},
\end{gather}
$A_{\pm}=A_x\pm i A_y$, $\mathbf{I}$ is the identity $4\times 4$ matrix. The Hamiltonian~(\ref{light-pair}) is written in a simplified form adapted to the pairs with small total momentum. It is seen that $H'(1,2)$ contains terms proportional to $A_{\pm}$ arising due to the hybridization of the $e$- and $h$-bands, in addition to usual dipole term $\mathbf{k}\mathbf{A}$. 

The decay rate is studied in the standard way making use of Fermi’s Golden rule with the perturbation $H'(1,2)$ for electron transitions from an initial state, in which there is an BEP in a state $|bs\rangle=\Psi_{bs}(\mathbf{r}_1-\mathbf{r}_2)$ and the electromagnetic field in the vacuum state $|\Omega\rangle$, to a final state, which involves two electrons in one of the possible band states $|f\rangle=\Phi_{s_1,\mathbf{k}_1,s_2,\mathbf{k}_2}^{\lambda_1,\lambda_2}$ and one photon with the wave vector $\mathbf{q}$ and the polarization $\mathbf{e}_{\nu}$. The total transition rate is obtained by summing over all possible final states: 
\begin{multline}
\Gamma =\frac{2\pi|M|}{\hbar} \sum_{\shortstack{$\scriptstyle \lambda_1,s_1$ \\ $\scriptstyle \lambda_2,s_2,\nu$}}\!\iiint \frac{d^3q}{(2\pi)^3} \frac{d^2k_1}{(2\pi)^2} \frac{d^2k_2}{(2\pi)^2} \\ \times \Bigl|\langle f|\otimes\langle \mathbf{q},\nu|H'(1,2)|bs\rangle\otimes|\Omega\rangle\Bigr|^2 \delta(\varepsilon_{bs}-\varepsilon_{\mathbf{k}_1,s_1;\mathbf{k}_2,s_2}^{\lambda_1,\lambda_2}-\varepsilon_q).
\label{rate}
\end{multline}

For simplicity, we assume here that the BEPs exist in an empty crystal, i.e. the $v$-band is not filled by electrons. This allows us to find the upper estimate of the decay rate, since it is clear that the filling of the bands leads to a decrease in the decay rate approximately as $f^{-2}$, with $f$ being the filling factors. However, reducing the number of unoccupied states in the bands is not the only effect produced by the presence of a large number of electrons. There are also such effects as screening of the \textit{e-e} interaction, electron correlations, etc. They require serious study, which is beyond the scope of this paper.

The decay rate is calculated directly from Eq.~(\ref{rate}), which is first simplified analytically taking into account the conservation of energy and momentum. A further simplification is made using the fact that the wave vector of the photon arising from the decay of the BEP is small. The final integration is carried out numerically. 

\begin{figure}[t]%
\centerline{\includegraphics*[width=1.\linewidth,height=0.6\linewidth]{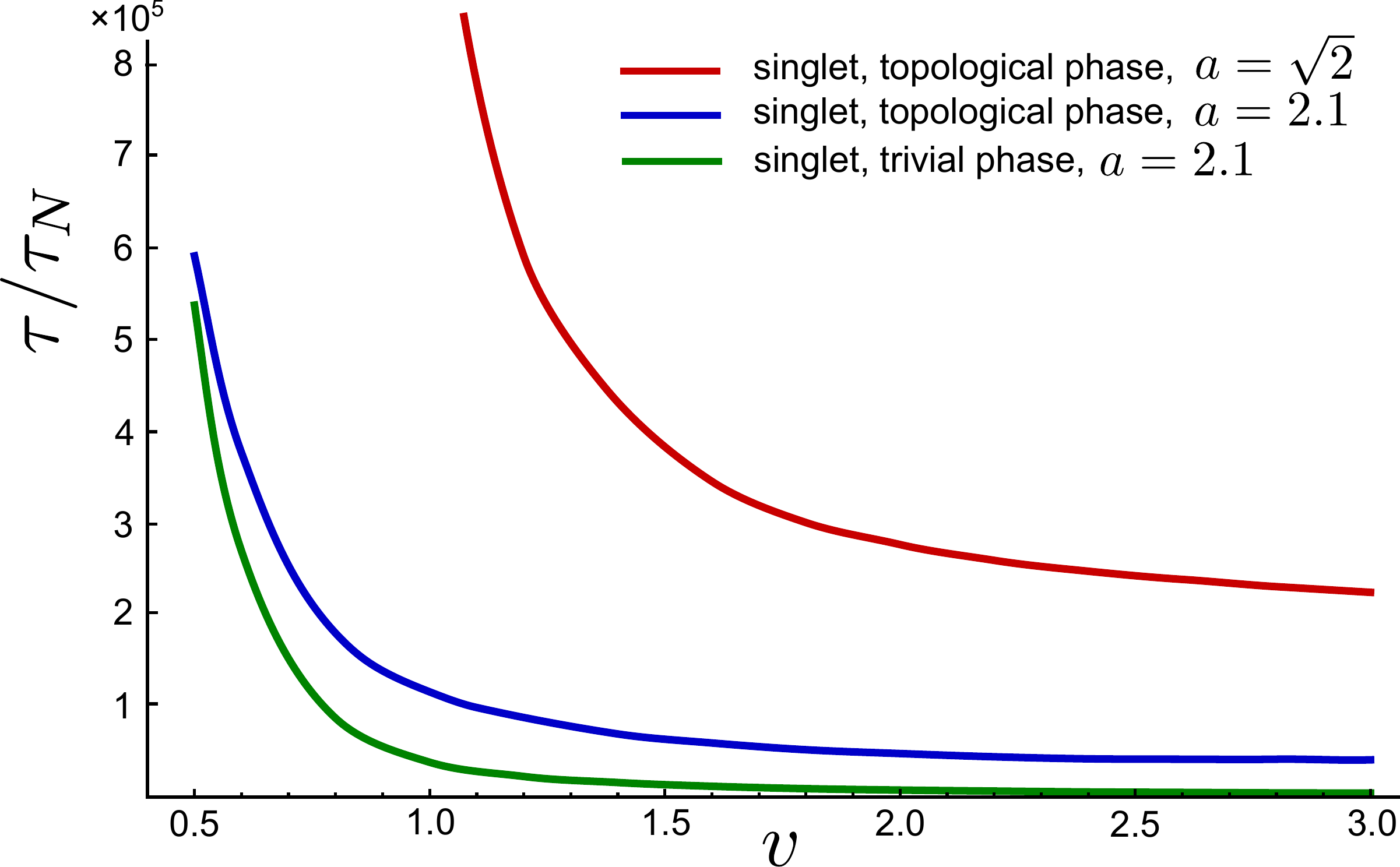}}%
\caption{The radiative decay time of the singlet BEPs as a function of the interaction potential for different types of the band structure. The calculations are made for $r_0=2,0$.} 
\label{fig4}
\end{figure}

The main results are presented in Fig.~\ref{fig4}, where the decay time is normalized to $\tau_N=\kappa \hbar/(4\pi^2e^2)(|B/M|)^{1/2}$. First of all, note, that the decay time turns out to be unexpectedly large compared with the radiative decay of excitons in direct-gap semiconductors. Numerical estimations using the parameters close to those of the heterostructures HgTe/CdHgTe give $\tau_N \approx 2\cdot 10^{-14}$~s and $a\approx 4$. In this case the decay time is estimated as $\tau \sim 10^{-9}$~s. 

Such a long decay time is primarily due to the structure of the two-electron wave functions in the initial and final states. First, the wave function of the initial and final states weakly overlap, since in the initial state the electrons are localized, and in the final state they are free. Second, the spinor components of both states have different signs, so some terms in their product are partially canceled. But more important is the restriction imposed by the energy and momentum conservation law on the phase volume, where the transition is possible. 

As for the spin, the spin state does not change in the decay process.

The BEPs break up mainly into the electrons in the $v$-band, i.e. through the $vv$-channel. The transitions through the $cv$-channel also occur when the energy $\varepsilon>0$, but their amplitude is relatively small. 

We have calculated the decay time for a variety of the model parameters and come to the following conclusions. The BEPs, which exist in the topological phase with nearly flat bands in the extrema at $a=\sqrt{2}$, have the largest decay time. In addition, they have the largest binding energy. In the topological phase with a quadratic spectrum in the extrema, which exists at $a>2$, the decay time is much shorter than in the above case. Nevertheless this time is much longer than that in the trivial phase, with other things being equal. These facts show that both the topological properties of the electronic states and the band dispersion near the extrema play an important role in stability of the BEPs.

We have explored also the triplet states and found that the rate of their decay through the $vv$-channel is extremely low, but not forbidden, in contrast the excitons. Radiative transitions in this channel become possible in the expansion of a higher order in the photon wave vector $\mathbf{q}$, so that the decay rate is small in the parameter $q^2|B/M|$. The transitions through the $cv$-channel have no such limitation, but this process is not so interesting.

\section{Conclusion}

We pay attention to the possible manifestation of the BEPs in transport properties and non-equilibrium processes in solids and especially in topologically nontrivial materials. In order to find out whether the BEP decay does not limit the possibility of realizing these effects in actual materials, we have studied the radiative decay of the BEPs in two-dimensional materials described by the BHZ model. We have focused on the BEPs with the highest binding energy, which are singlet states, and found that the radiative decay time is rather large on the scale of the characteristic relaxation times of the electron system. For realistic conditions of the HgTe/CdHgTe heterostructures, the decay time is estimated at the nanosecond level; however, taking into account the filling of the band states one can expect that the decay time will even longer.

The relatively long decay time is mainly due to two factors that significantly distinguish the decay of BEPs from the decay of excitons: weak overlapping of the wave function of the free electrons, arising from the decay, and the wave function of the electrons bound in a pair, and the restriction imposed by the requirement of the energy and momentum conservation on the phase space where the radiative transition is possible.

The radiative decay rate strongly depends on the topological properties of the band states and the band dispersion. In the topological phase, the decay time is essentially longer than in the trivial one. This is because the two-particle wave functions have very different spinor structure in these phases. In the trivial phase, the strongly dominant component of the spinor is the one that corresponds to the configuration in which both electrons are in the hole band. In this configuration, the matrix element of the transition of electrons to the valence band is relatively large. In contrast, in the topological phase, all spinor components are close in order of magnitude. 

Within the BHZ model, the decay time significantly depends on the band hybridization parameter $a$, which determines the band dispersion. 
A feature of the topological phase is the possibility to realize a nearly flat band dispersion, which occurs at $a=\sqrt{2}$. In this case the longest decay time of the BEPs is reached. Given the fact that in this case the BEPs has the largest binding energy, one can expect that this type of the BEPs is the most stable one. 

Of course, the final conclusion about the decay time of the BEPs requires further study of other mechanisms of nonradiative decay and many-particle effects, but the presented results inspire optimism regarding the possible manifestations of the BEPs in nonequilibrium and collective processes in topologically nontrivial materials.

\textit{Acknowledgments.}---This work was supported by Russian Science Foundation (Grant No~16–12–10335).

\bibliography{decay}

\end{document}